\documentclass{aa}

\usepackage{graphicx}
\usepackage{txfonts}
\usepackage{siunitx}
\usepackage{arydshln}

\usepackage{CJKutf8}
\newcommand{\CNnames}[1]{{\begin{CJK}{UTF8}{gbsn}~(#1)~\end{CJK}}}
\newcommand{\methu}[0]{HD~140283\xspace}
\newcommand{\numax}[0]{\mbox{$\nu_\mathrm{max}$}\xspace}
\newcommand{\gaia}{\textit{Gaia}\xspace}

\begin{document}

   \title{Asteroseismic investigation of HD~140283: The Methuselah star}

   \author{M. S. Lundkvist\inst{1}
          \and
          J. R. Larsen\inst{1}
          \and
          Y. Li\CNnames{李亚光}\inst{2}
          \and
          M. L. Winther\inst{1}
          \and
          T. R. Bedding\inst{3}
          \and
          H. Kjeldsen\inst{1}
          \and
          T. R. White\inst{4,3}
          \and
          M. B. Nielsen\inst{5}
          \and
          G. Buldgen\inst{6}
          \and
          C. Guillaume\inst{6}
          \and
          A. L. Stokholm\inst{5}
          \and
          D. Huber\inst{2,3}
          \and
          J. L. R{\o}rsted\inst{1}
          \and
          P. Mani\inst{3}
          \and
          F. Grundahl\inst{1}
          }

   \institute{Stellar Astrophysics Centre, Department of Physics and Astronomy, Aarhus University, 8000 Aarhus C, Denmark\\
              \email{lundkvist@phys.au.dk}
         \and
             Institute for Astronomy, University of Hawai`i, Honolulu, HI 96822, USA
         \and
             Sydney Institute for Astronomy, School of Physics, University of Sydney NSW 2006, Australia 
         \and
             Sydney Informatics Hub, Core Research Facilities, University of Sydney, NSW 2006, Australia
         \and
             School of Physics and Astronomy, University of Birmingham, Birmingham B15 2TT, UK
         \and
             STAR Institute, Universit{\'e} de Li{\`e}ge, Li{\`e}ge, Belgium
             }

   \date{Received: July 7 2025; accepted September 30, 2025}
 
  \abstract 
   {\methu\ is a well-studied metal-poor subgiant and a Gaia benchmark star, often used for testing stellar models due to its proximity, brightness, and low metallicity ($[\mathrm{Fe}/\mathrm{H}] = -2.3$ dex).}
   {Here we present the first asteroseismic analysis of \methu, providing improved constraints on its fundamental properties.}
   {The star was observed by TESS in 20-second cadence during Sector 51. We extracted a custom light curve and performed a frequency analysis, revealing a rich spectrum of solar-like oscillations including mixed modes. These were combined with parameters from the literature to provide constraints on our model inference performed with BASTA.}
   {Using a dense grid of models, we find a mass of $0.75 \pm 0.01 \ \mathrm{M}_\odot$, a radius of $2.078 \substack{+0.012\\-0.011} \ \mathrm{R}_\odot$, and an age of $14.2 \pm 0.4\ \mathrm{Gyr}$, in agreement with the upper limit set by the age of the Universe within $1\sigma$. The observed frequency of maximum power, $\left(\nu_\mathrm{max}\right)_\mathrm{obs} = 611.3 \pm 7.4 \ \mu\mathrm{Hz}$, is significantly higher than predicted from standard scaling relations ($\left(\nu_\mathrm{max}\right)_\mathrm{mod} = 537.2 \substack{+2.9\\-1.8} \ \mu\mathrm{Hz}$), extending known deviations into the metal-poor regime.}
   {To our knowledge, the oscillations in \methu have the highest $\nu_\mathrm{max}$ of any metal-poor star to date, which will help to advance our understanding of oscillations in metal-poor stars in general. The results demonstrate the value of asteroseismology for precise age determination in old halo stars and taking custom abundances and opacities into account during the modelling is probably important for further improving models of such stars. In addition, a detailed characterisation of metal-poor stars, such as HD~140283, will also help advance our understanding of Population III stars and their impact on future stellar generations.}


   \maketitle

\section{Introduction}
\label{sec:intro}

Unravelling the history of the Milky Way requires precise knowledge of its oldest stellar populations \citep[e.g.][]{ref:freeman2002}. Metal-poor stars, as relics from the early Universe, provide important clues about the formation and chemical enrichment history of the Galaxy.  
Among the various techniques for characterising stars, asteroseismology---the study of stellar oscillations---offers an unparalleled opportunity to determine precise stellar ages, particularly for old, low-mass stars where traditional methods suffer from degeneracies \citep[see, e.g.][]{ref:soderblom2010, ref:chaplin2013, ref:lebreton2014}. Recently, the power of asteroseismology has been applied to measuring precise masses and ages for several metal-poor stars \citep{Bedding2006, ref:deheuvels2012, ref:valentini2019, ref:chaplin2020, ref:huber2024, Larsen25, ref:lindsay2025, ref:marasco2025}. 

One of the most well-studied metal-poor stars is \methu, also known as the `Methuselah’ star. 
With a metallicity of $[\mathrm{Fe}/\mathrm{H}] \sim -2.3$ (see Sec.~\ref{sec:existingdata}) and a brightness that places it among the most accessible metal-poor halo stars \citep{ref:creevey2015}, \methu is an excellent benchmark for testing stellar models and is one of the benchmark stars for the \gaia mission. Its proximity and extensive characterisation---including Gaia parallax \citep{ref:gaiadr3_2023} and interferometric radius measurements \citep{ref:creevey2015, ref:karovicova2018, ref:karovicova2020}---make it a cornerstone in the study of metal-poor subgiants.

\begin{figure}
   \centering
   \includegraphics[width=\hsize]{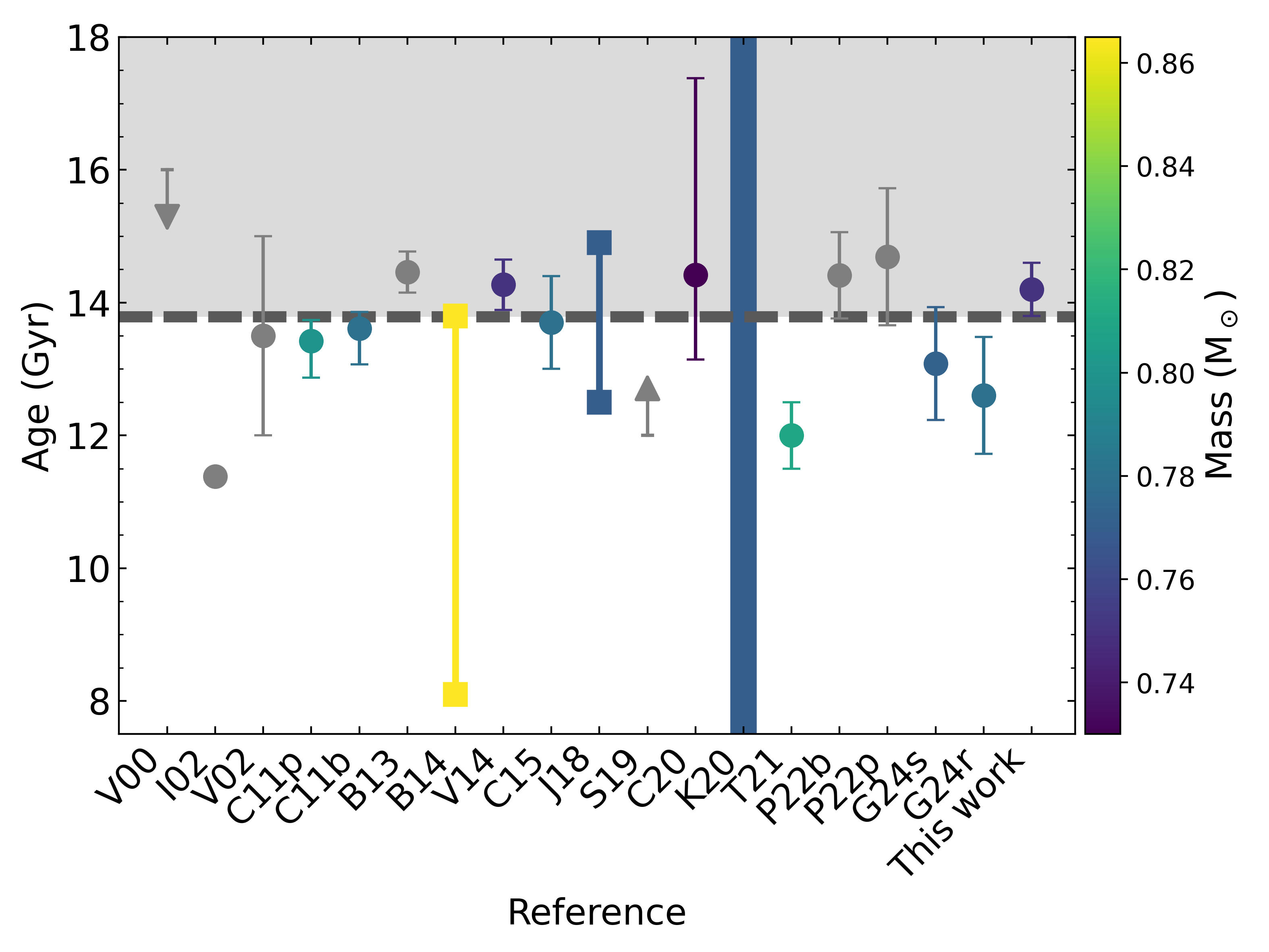}
      \caption{Compilation of literature ages colour-coded by mass for \methu (the result of this work is also given). The circles show age estimates, the squares give age ranges, and the 
      arrows are upper and lower limits. The dashed line at $13.787 \ \mathrm{Gyr}$ indicates the age of the Universe \citep{ref:planck2020} and its uncertainty. A table of the references can be found appendix~\ref{sec:app_ages}.
              }
    \label{fig:age_lit}
\end{figure}

The age of \methu has been estimated in several works using various techniques. Several studies focussed on examining the role of the mixing-length parameter \citep{ref:creevey2015, ref:joyce2018, ref:tang2021, ref:guillaume2024} and/or taking the abundances into account \citep{ref:bond2013, ref:vandenberg2014, ref:sahlholdt2019, ref:karovicova2020, ref:guillaume2024}. Some of these also varied the atmospheric boundary conditions \citep{ref:joyce2018, ref:tang2021}, while others focussed on the reddening or extinction \citep{ref:creevey2015, ref:plotnikova2022}. These studies leave no doubt that the star is old, but the modelling choices have resulted in a spread of age estimates, sometimes placing \methu slightly older than the Universe itself (see Fig.~\ref{fig:age_lit}). This highlights the limitations of classical methods and the need for asteroseismic constraints to break the degeneracies that exist between several of these parameters \citep[see, e.g.][]{ref:creevey2015}.

Asteroseismology can, not only provide ages precise to $<20\%$ \citep{ref:tayar2022} for main-sequence and sub-giant stars, but also precise masses. As is evident from Fig.~\ref{fig:age_lit}, there is a considerable spread between the existing mass estimates for \methu, with an overall correlation between higher masses and younger ages, as to be expected.

However, despite being an ideal target, an asteroseismic analysis of \methu has not been carried out until now. In this work, we present the first detection of solar-like oscillations in \methu using data from the TESS mission. We identify both radial and dipolar mixed modes, allowing detailed modelling of the star's internal structure and a precise age determination using the BASTA framework \citep{Aguirre22}.

\begin{figure}
   \centering
   \includegraphics[width=\hsize]{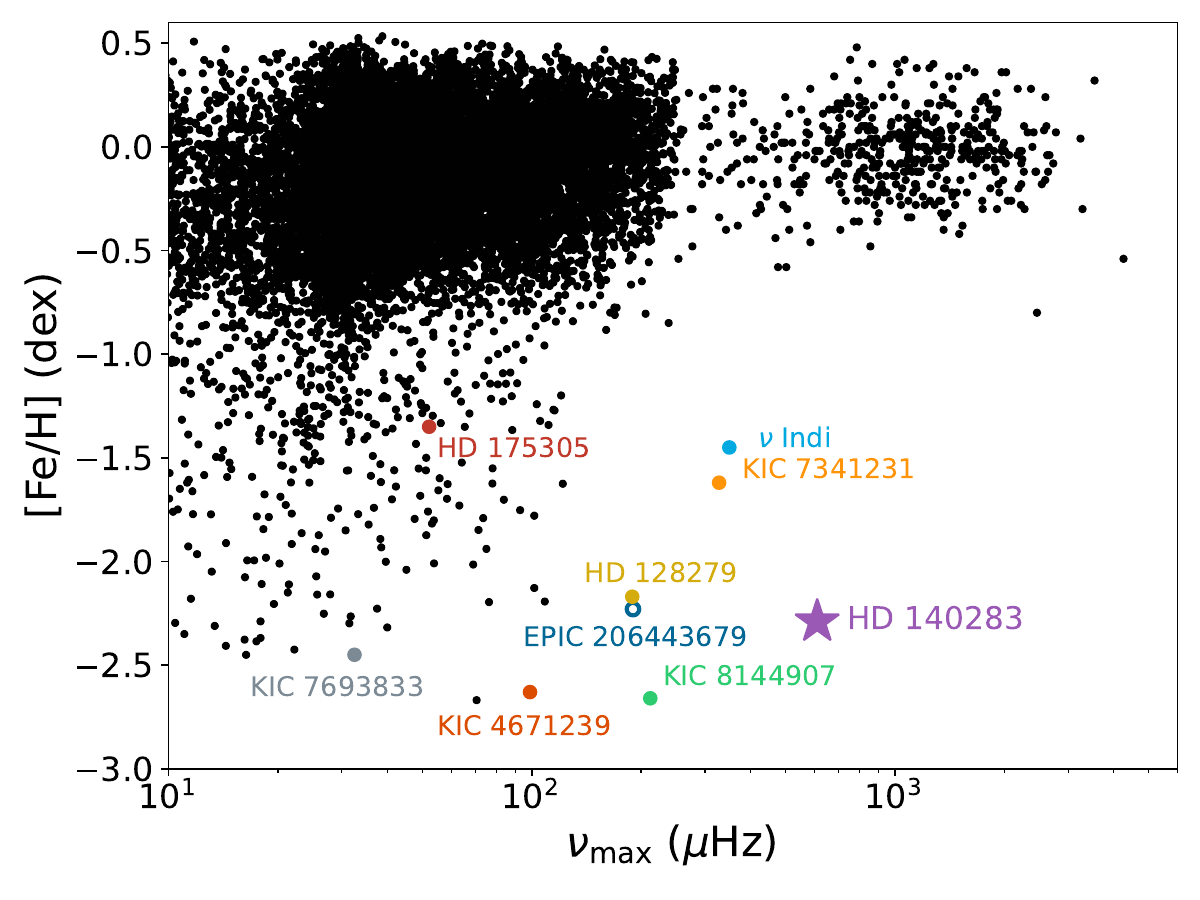}
      \caption{Iron abundance vs frequency of maximum power for stars with asteroseismic detections, modified from a similar plot by \citet{ref:huber2024}. The star shows \methu (this work), while the points are literature values. The black points are from \citet{ref:pinsonneault2014}, \citet{ref:serenelli2017}, \citet{ref:metsuno2021} and \citet{ref:stasik2024}. The filled coloured points show stars with modelling of individual frequencies; $\nu$~Indi \citep{Bedding2006, ref:chaplin2020}, KIC~7341231 \citep{ref:deheuvels2012}, KIC~7693833 and KIC~4671239 \citep{Larsen25}, KIC 8144907 \citep{ref:huber2024}, and HD 175305 and HD 128279 \citep{ref:lindsay2025}, while the open circle shows EPIC 206443679 \citep{ref:valentini2019} with modelling of the global seismic parameters.}
    \label{fig:feh_numax}
\end{figure}

To place \methu in context, it is instructive to compare it with other known metal-poor stars with a seismic detection in a plot of $[\mathrm{Fe}/\mathrm{H}]$ as a function of the frequency of maximum power, \numax (a proxy for the evolutionary stage). This plot can be seen in Fig.~\ref{fig:feh_numax} and reveals the unique position of \methu in terms of metallicity and the evolutionary stage. Here, it is evident that \methu is distinct in this sample, with a significantly higher \numax than any previously studied metal-poor star, making it an important data point for understanding the behaviour of oscillations in the metal-poor regime.

This paper is structured as follows: in Sec.~\ref{sec:existingdata} we summarise the known properties of \methu, while Sec.~\ref{sec:tessdata} describes the TESS observations. The construction of the power spectrum and extraction of oscillation frequencies are presented in Sec.~\ref{sec:ps}, while the modelling is the subject of Sec.~\ref{sec:modelling}. In Sec.~\ref{sec:results} we describe the results of the analysis and the impact of including different constraints in the model inference. Sec.~\ref{sec:discussion} explores the implications for the \numax scaling relations and the Galactic origin of \methu, before we present our conclusions in Sec.~\ref{sec:conclusion}.


\section{Properties of \methu}
\label{sec:existingdata}

\begin{table}
\caption{Adopted parameters of \methu.}
\label{tab:knownparams}
\centering                        
\begin{tabular}{l c c}      
\hline\hline                
Parameter & Value  & Reference \\ 
\hline                     
$T_\mathrm{eff}$            & $5792 \pm 55$ K                       & 1 \\
$[\mathrm{Fe}/\mathrm{H}]$  & $-2.29 \pm 0.10 \pm 0.04$ dex                   & 1 \\
$\theta_\mathrm{LD}$ & $0.325 \pm 0.006$ mas & 1 \\
$R_\mathrm{int}$            & $2.142 \pm 0.040 \ \mathrm{R}_\odot$\tablefootmark{a}  & \\
$G_\mathrm{BP}$    & $7.321$                              & 2, 3 \\
$G_\mathrm{RP}$    & $6.562$                              & 2, 3 \\
$G$                         & $7.036$                              & 2, 3 \\
$\varpi$                    & $16.305 \pm 0.026$ mas\tablefootmark{b} & 3 \\
\hline
\end{tabular}
\tablefoot{
1: \citet{ref:karovicova2020}
2: \citet{ref:riello2021}
3: \citet{ref:gaiadr3_2023}
\tablefoottext{a}{Derived using $\theta_\mathrm{LD}$ from 1 and $\varpi$.}
\tablefoottext{b}{Zero-point corrected parallax, see Sect.~\ref{subsec:modelling}.}
}
\end{table}

\methu has been observed extensively by both ground-based and space-based facilities, and it is the second most metal-poor star in the Gaia FGK star catalogue \citep{ref:soubiran2024}.\footnote{The most metal-poor star in that catalogue is the luminous red giant HD~122563 \citep{ref:creevey2019}.} It has been observed three times with CHARA, where \citet{ref:karovicova2020} most recently found an interferometric radius of $R_\mathrm{int} = 2.167 \pm 0.041 \ \mathrm{R}_\odot$, in between those reported earlier \citep{ref:creevey2015, ref:karovicova2018}. Here, we combined the angular diameter determined by \citet{ref:karovicova2020}, $0.325\pm0.006$ mas, with the updated parallax from Gaia DR3 to re-derive the interferometric radius, which we adopted (see Table~\ref{tab:knownparams}). 

Several literature values for the metallicity ($[\mathrm{Fe}/\mathrm{H}]$) of \methu exist \citep[see, e.g.][]{ref:schuster1989, ref:vandenberg2000, ref:nissen2002, ref:bond2013, ref:bensby2014, ref:jofre2015, ref:adibekyan2020, ref:amarsi2022, ref:liezzeddine2023}. Here, for consistency, we used the value $[\mathrm{Fe}/\mathrm{H}] = -2.29 \pm 0.10 \pm 0.04$ dex found by \citet{ref:karovicova2020}, which is in agreement with, for instance, the recent value of $-2.28 \pm 0.02$ found by \citet{ref:amarsi2022} based on 3D non-LTE analyses. We also used the effective temperature ($T_\mathrm{eff}$) from \citet{ref:karovicova2020}.

The relative composition of \methu differs from that of the Sun \citep[see, e.g.][]{ref:guillaume2024}, notably it is enhanced in $\alpha$-elements and in particular oxygen \citep{ref:nissen2002, ref:bond2013, ref:vandenberg2014, ref:siqueira2015}. The $\alpha$-enhancement values for \methu differ between different studies \citep{ref:bensby2014, ref:vandenberg2014, ref:jofre2015, ref:siqueira2015, ref:chen2020, ref:garcia2021, ref:spite2022, ref:casamiquela2025} with a median value of $0.3$~dex; where $[\alpha/\mathrm{Fe}]$ was not listed in the paper we computed it from the available abundances of magnesium, calcium, silicium, and titanium. 

In the modelling (Sec.~\ref{sec:modelling}), we included the Gaia parallax ($\varpi$) and the \textit{Gaia} $G$, $G_\mathrm{BP}$, and $G_\mathrm{RP}$ magnitudes, which are available for \methu in Gaia DR3 \citep{ref:gaiadr3_2023} and can be found in Table~\ref{tab:knownparams}. 

The reddening of \methu is small. It was found by \citet{ref:melendez2010} to be $E(B-V) = 0.000 ± 0.002$ while \citet{ref:vandenberg2014} found a value of $E(B-V) = 0.004$. Thus, here we neglected reddening following, for example, \citet{ref:karovicova2020}.


To check for close companions that might contaminate the TESS light curve, we first used the Gaia data releases 1--3 \citep{ref:gaiadr1_2016, ref:gaia2016, ref:gaiadr2_2018, ref:gaiadr3_2023}). Through the Gaia archive, we queried a $20''$ circle around the coordinates of \methu, exceeding the TESS pixel scale of $21''$ \citep{ref:pelaez2024}, and the only source to show up was \methu\ itself. Given the high proper motion of \methu\ of ${\sim}1.2 \ \mathrm{arcsec} \cdot \mathrm{yr}^{-1}$ (Gaia DR3 values are $\mu_{\alpha *} = -1115.141 \ \mathrm{mas} \cdot \mathrm{yr}^{-1}$ and $\mu_\delta = -303.573 \ \mathrm{mas} \cdot \mathrm{yr}^{-1}$)\footnote{Definition of proper motion components can be found at \url{https://gea.esac.esa.int/archive/documentation/GDR3/Gaia_archive/chap_datamodel/sec_dm_main_source_catalogue/ssec_dm_gaia_source.html#gaia_source-ref_epoch}.}, a chance alignment would have been revealed. Furthermore, the \gaia RUWE value for \methu is 1.064, consistent with a single star \citep{Castro-Ginard2024}.
We also checked for bound companions in the available speckle imaging from Gemini\footnote{\url{https://exofop.ipac.caltech.edu/tess/target.php?id=290565106}} \citep{ref:howell2021}, which excluded a bound companion that would be prominent in the power spectrum beyond $0.1''$. We conclude that the TESS light curve is unlikely to be contaminated by flux from neighbouring stars.


\section{TESS time series}
\label{sec:tessdata}

\begin{figure}
   \centering
   \includegraphics[width=\hsize]{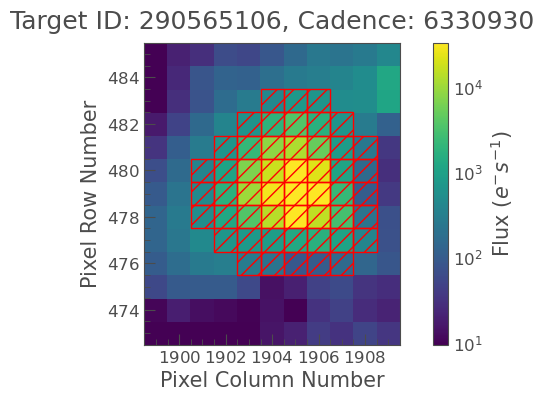}
      \caption{Example of a target pixel file for \methu. Our custom mask is shown in red.
              }
      \label{fig:tpf}
\end{figure}

\begin{figure*}
   \resizebox{\hsize}{!}{\includegraphics{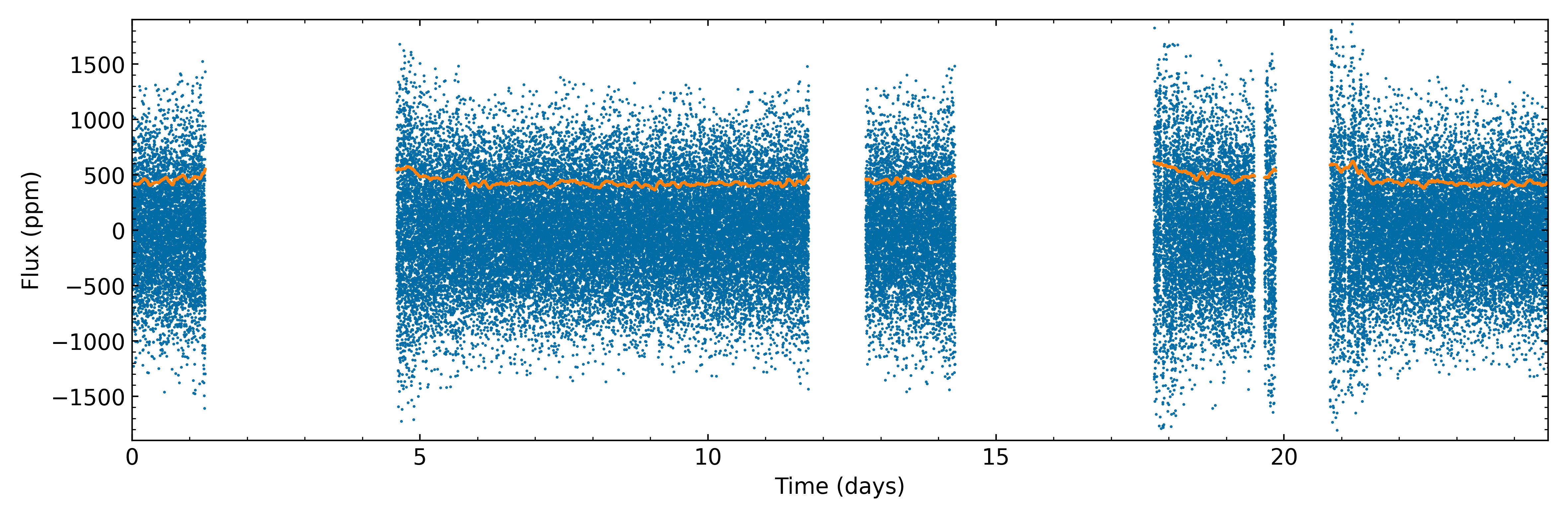}}
      \caption{Time series of \methu. The flux is shown in blue and the measurement uncertainties in orange. The gaps are explained in the text.
              }
      \label{fig:timeseries}
\end{figure*}

\methu\ was observed by TESS in 20-sec cadence in Camera 1 during Sector 51 (we note, that it was recently observed again in Sector 91, which we have not included in the analysis). The Earth crossed Camera 1 at the start of both orbits, saturating the detectors and causing strong glints and scattered light signals. The Moon also crossed Camera 1 in the second orbit. As a result, there are large gaps in the observations \citep{TESS-DRN-51}. The light curves produced by the TESS Science Processing Operations Centre \citep[SPOC,][]{Jenkins2016} tend to be conservative in flagging sections of the light curve that may be affected by scattered light. Additionally, the SPOC light curve of \methu\ showed segments that were affected by increased scatter. We therefore used \texttt{lightkurve} \citep{Lightkurve2018} to construct a custom light curve from the target pixel files.

We used simple aperture photometry with a custom aperture, which is shown in Fig.~\ref{fig:tpf}, to extract the light curve. The aperture was chosen by first selecting pixels with a median flux greater than three times the standard deviation about the overall median, and then widening this selection by one pixel in each direction. We excluded cadences where the quality flags identified Argabrightening and cosmic ray events. Subsequently, we used principal component analysis to determine the five most significant trends in pixels outside of the target aperture. We used linear regression to detrend our raw light curve against these principal components, resulting in our corrected light curve.

Finally, we estimated the measurement uncertainties from the scatter in the time series following equations 5 and 6 of \citet{ref:kjeldsen2025} using a half-width of $0.05$~days. Fig.~\ref{fig:timeseries} shows the corrected time series and the estimated measurement uncertainties.


\section{Measurement of oscillation frequencies}
\subsection{Calculating the power spectrum}
\label{sec:ps}

   \begin{figure}
   \centering
   \includegraphics[width=\hsize]{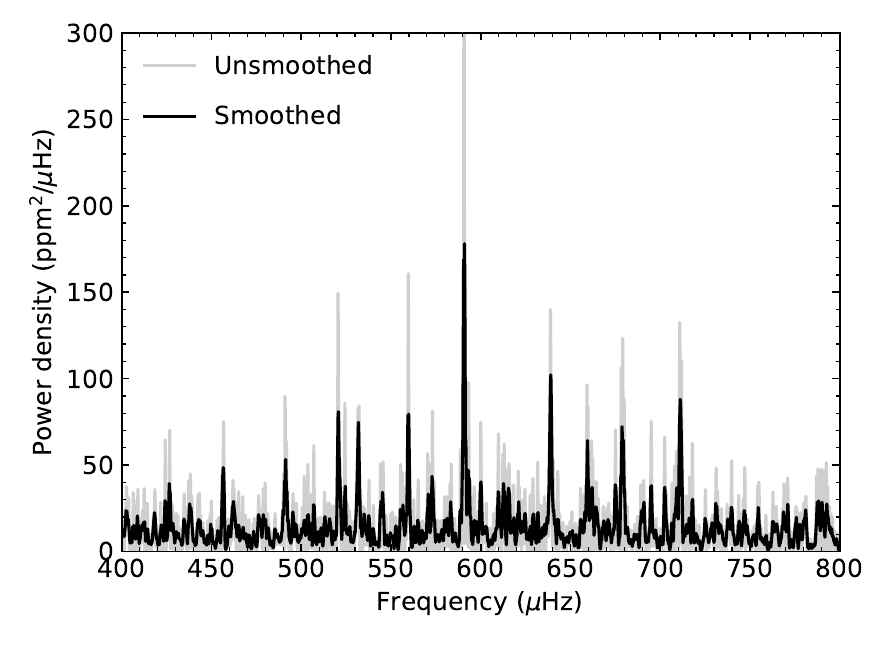}
      \caption{Power spectrum of \methu: unsmoothed (grey) and smoothed to 1~$\mu$Hz (black) with the oscillation frequencies clearly seen. The measured frequency of maximum oscillation power is $\nu_\mathrm{max} = 611.3 \pm 7.4 \ \mu\mathrm{Hz}$.
              }
      \label{fig:pds}
   \end{figure}

We calculated the weighted power spectrum using a standard sine-wave fitting technique \citep[see, e.g.][]{ref:kjeldsen1992, ref:frandsen1995, ref:handberg2014}. The squared inverse of the measurement uncertainties were used as the weights. The resulting power spectrum of the oscillations can be seen in Fig.~\ref{fig:pds}. When computing the power spectrum, we tested the effect of removing additional data points from the time series if they had a flag indicating a possible decreased quality. However, as we were computing a weighted power spectrum, using the full time series resulted in the lowest noise level and the best spectral window.

The comb-like structure characteristic of solar-like oscillations is immediately visible from Fig.~\ref{fig:pds}. Such solar-like (p-mode) oscillations of high radial order ($n$) and low spherical degree ($\ell$) are approximately described using the asymptotic relation \citep{ref:tassoul1980}:
\begin{equation}
    \label{eq:asymptoticrel}
    \nu_{n,\ell} \approx \Delta\nu \left( n + \frac{\ell}{2} + \epsilon \right) - \delta\nu_{0,\ell} \ .
\end{equation}
Here, $\Delta\nu$ is the large frequency separation between modes of identical degree and consecutive radial order, giving rise to the overall regularity of the p-modes in the power spectrum. The quantity $\epsilon$ is a dimensionless offset of order unity. The small separation, $\delta\nu_{0,\ell}$, is particularly sensitive to the stellar age in main-sequence stars because it probes variations in the gradient of the sound speed close to the core \citep{Christensen-Dalsgaard1988}. However, for sub-giants it was shown by \citet{ref:white2011} that the small separation loses some of its age diagnostic potential.

As can be seen from the power spectrum in Fig.~\ref{fig:pds}, the power excess is centred around $600 \ \mu$Hz. Using the pySYD pipeline \citep{ref:huber2009,ref:chontos2022} we determined the frequency of maximum power to be $\numax = 611.3 \pm 7.4 \ \mu$Hz. This means that \methu has, to our knowledge, the highest measured value of $\nu_\mathrm{max}$ for any metal-poor star ($[\mathrm{Fe}/\mathrm{H}] < -1$; see also Fig.~\ref{fig:feh_numax}).

To identify the $\ell=0$ modes and determine the value of $\Delta\nu$ we constructed {\'e}chelle diagrams \citep{ref:grec1983, Hey2022}. The optimal $\Delta\nu$ value should align the ridge formed by the $\ell=0$ modes vertically, and we estimated $\Delta\nu=39.5$ $\mu$Hz (the {\'e}chelle diagram is shown in Fig.~\ref{fig:echelle}). 

The {\'e}chelle diagram showed additional peaks that we identified as mixed modes of angular degree $\ell=1$. This is expected for a subgiant star such as \methu, and they are discussed further in Sec.~\ref{subsec:freqs}. For an {\'e}chelle diagram of a star in a very similar evolutionary stage, the reader is referred to KIC 8702606 ($\Delta\nu = 39.4 \ \mu\textrm{Hz}$) in Fig.~B1 of \citet{ref:t_li2020}. Interestingly, most of the $l=2$ modes in \methu were too weak to detect reliably. 


\subsection{Extraction of frequencies}
\label{subsec:freqs}

Frequencies were extracted by three different methods, namely, the one described below, PBJam \citep{ref:nielsen2021}, and the method employed by \citet{ref:kjeldsen2025}, where the most significant peaks are extracted from the CLEANed power spectrum. Here, we report the frequencies returned by the method detailed below because it returned the greatest number of modes that agreed with at least one of the other methods. These modes are listed in Table~\ref{tab:freqs}, with the addition of a single $\ell=2$ mode, which was returned by the other two methods. It was not used as an input in the final modelling (see Sec.~\ref{sec:modelling}), but we verified that the best-fitting model was consistent with this $\ell=2$ mode and tested its effect on the reported uncertainties (see Sec.~\ref{subsec:other_constraints}). It is worth noting that all three methods agreed on the six observed radial modes.

Next, we identified the positions of $\ell = 1$ mixed modes using a combination of stretched-period {\'e}chelle \citep{Gaulme2022,Ong2023} and stretched-frequency {\'e}chelle \citep{ref:li2024}.
The mixed modes result from the coupling of g-modes in the core and p-modes in the envelope. For this star, we expect both types of modes to be in the asymptotic regime \citep{Unno1989,Mosser2018,Ong2023}, where g-modes and p-modes are approximately evenly spaced in period and frequency, respectively. Mixed modes deviate from these regular-spacing patterns but can be recovered by constructing the stretched-period and stretched-frequency {\'e}chelle diagrams. The modes should be vertically aligned in these {\'e}chelles if the correct asymptotic parameters are used. The asymptotic parameters we found to optimise the stretched diagrams are $\Delta\Pi=147$ s (period spacing), $q=0.41$ (coupling strength), $\epsilon_p = 0.18$, and $\epsilon_g = 0.65$. The high coupling strength is consistent with the findings by \citet[][see their Fig.~6]{ref:mosser2017}, especially considering the fact that the star is metal-poor which, at least for red giants, correlates with a higher value of $q$ \citep{ref:kuszlewicz2023}.

\begin{figure*}
   \resizebox{\hsize}{!}{\includegraphics{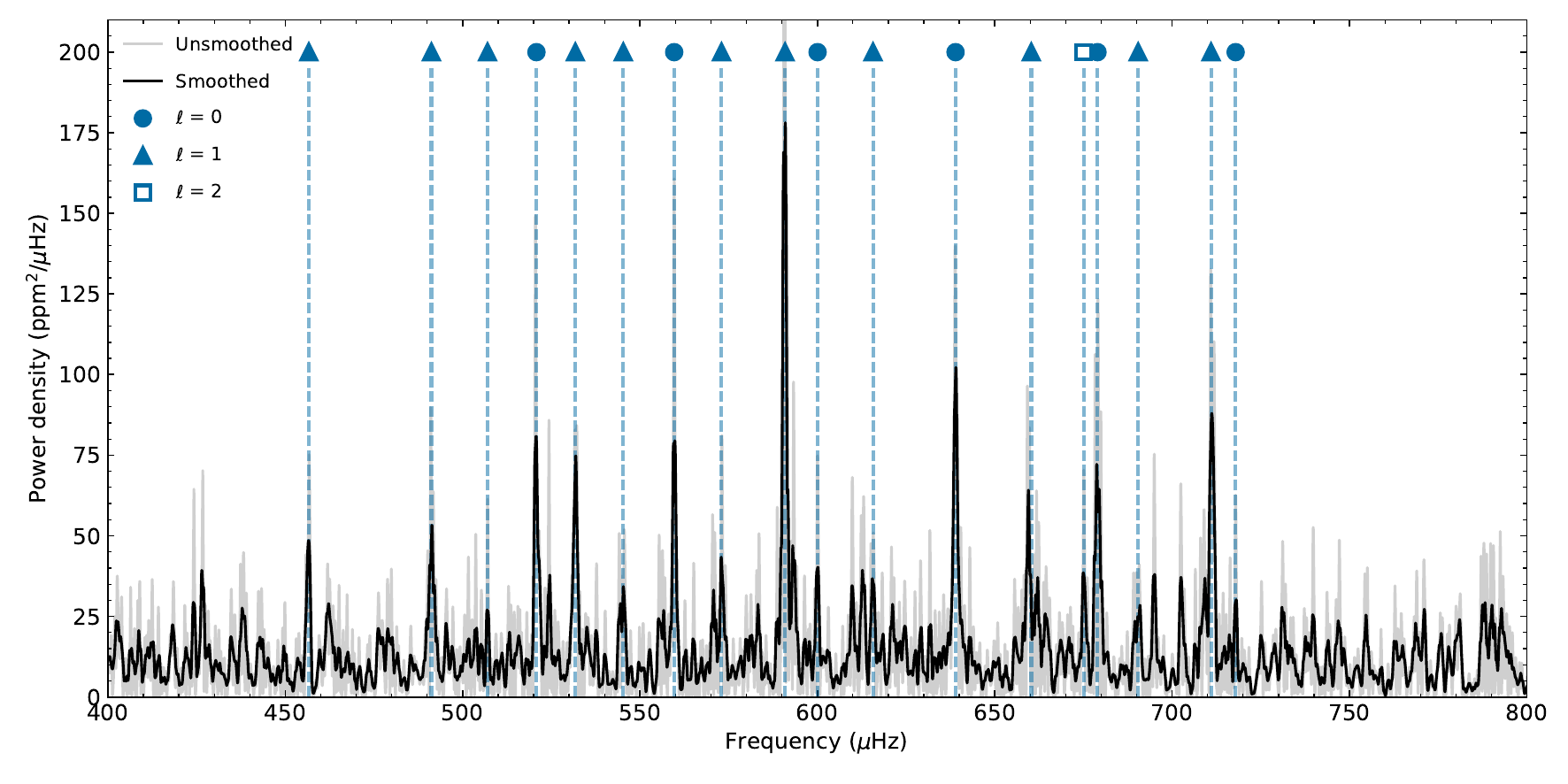}}
      \caption{Power spectrum of HD~140283: un-smoothed (grey) and smoothed to 1~$\mu$Hz (black) with the identified frequencies overplotted (see Table~\ref{tab:freqs}). The radial modes are indicated by circles, the dipolar modes by triangles, and the single quadrupolar mode that is not used in the reported modelling is shown with the open square.
              }
      \label{fig:pds_freqs}
\end{figure*}

Based on the initial identifications, we compiled a list of frequencies to use as initial guesses for fitting the power spectrum, in the process known as 'peakbagging' \citep{Handberg2011,Davies2016}. We followed the procedure described by \citet{LiY2020}. Each oscillation mode was modelled as a Lorentzian profile, and rotational splittings were neglected due to the relatively short duration of the time series. From the fitting of the radial modes we obtain $\Delta\nu = 39.468 \pm 0.051 \ \mu$Hz. The final frequencies can be seen overplotted on the power spectrum in Fig.~\ref{fig:pds_freqs} and are listed in Table~\ref{tab:freqs}.

\begin{table}
\caption{Extracted frequencies along with their spherical degree.}
\label{tab:freqs}
\centering
\renewcommand{\arraystretch}{1.2}
\begin{tabular}{c c}
\hline\hline
Frequency ($\mu$Hz) & Degree ($\ell$) \\
\hline
    $456.64 \pm 0.25$ & 1 \\
    $491.21 \pm 0.39$ & 1 \\
    $507.03 \pm 0.15$ & 1 \\
    $520.81 \pm 0.32$ & 0 \\
    $531.78 \pm 0.30$ & 1 \\
    $545.30 \pm 0.63$ & 1 \\
    $559.63 \pm 0.21$ & 0 \\
    $572.98 \pm 0.74$ & 1 \\
    $590.89 \pm 0.29$ & 1 \\
    $600.06 \pm 0.20$ & 0 \\
    $615.69 \pm 0.37$ & 1 \\
    $638.95 \pm 0.31$ & 0 \\
    $660.29 \pm 0.93$ & 1 \\
    $678.96 \pm 0.42$ & 0 \\
    $690.43 \pm 1.56$ & 1 \\
    $710.98 \pm 0.41$ & 1 \\
    $717.90 \pm 0.15$ & 0 \\
    \hdashline 
    $675.05 \pm 1.10$ & 2 \\
\hline                              
\end{tabular}
\tablefoot{The $\ell=2$ mode was not used for producing the final modelling results presented in Table~\ref{tab:modparams}. The modes have not been corrected for the Doppler shift of the star.}
\end{table}

\section{Modelling HD~140283 with BASTA}
\label{sec:modelling}

We employed the modelling approach in which the observed parameters are used to infer the stellar evolutionary models that best represent the star across a large grid of models. This entails creating specialised grids tailored to the suitable parameter space \citep{Stokholm19, Verma2022, Winther23, Larsen25}. This ensures that the stellar models and the tracks they produce yield sufficiently high sampling in the region of the parameter space where the likely solutions exist. Furthermore, various configurations of stellar properties can yield suitable solutions to the observed data due to the degeneracies of stellar modelling \citep{Basu17} -- a complexity that the grid modelling approach is well-equipped to evaluate.

\subsection{The model grid built for HD~140283}
\label{subsec:modelgrid}

\begin{table}[]
\centering
\renewcommand{\arraystretch}{1.1}
\caption{Variable parameters of the tailored grid for \methu with the lower and upper bounds indicated.
}
\begin{tabular}{lcc}
\hline
\multicolumn{1}{c}{Stellar parameter} & \multicolumn{1}{c}{Lower bound} & \multicolumn{1}{c}{Upper bound} \\ \hline 
$M$ $(\mathrm{M}_\odot)$                        &  0.70               & 0.85                       \\
$[\mathrm{Fe}/\mathrm{H}]_\mathrm{ini}$ (dex)   &  $-2.6$             & $-1.9$                     \\
$[\alpha/\mathrm{Fe}]_\mathrm{ini}$ (dex)       &  0.2                & 0.4                        \\
$\alpha_\mathrm{MLT}$                           &  1.70               & 1.90                       \\
$Y_\mathrm{ini}$ (dex)                          &  0.245              & 0.260                      \\
$\Delta\nu$ ($\mu$Hz)                           &  35                 & 45                         \\
\hline
\end{tabular}
\tablefoot{Apart from $\Delta\nu$ which constrains the temporal bounds of the grid, the remaining parameters are the initial parameters of the tracks, which have been uniformly sampled.}
\label{tab:GridParams}
\end{table}

The GARSTEC stellar evolution code \citep{Achim08} was used to calculate the stellar models in the grid . The construction of the grid utilised a quasi-random Sobol sampling \citep{Sobol67} to uniformly distribute the tracks of stellar evolution across the parameter space of sampled initial parameters (see Table~\ref{tab:GridParams}). For details on the specifics on equations of state, opacities, nuclear reaction rates, and $\alpha$-enhanced element abundances see Sec.~4 of \citet{Larsen25}. It should be noted that $[\alpha/\mathrm{Fe}]$ was only sampled in steps of $0.1$~dex because the opacity tables are only available in GARSTEC at this interval.

Our tailored grid for \methu is one of the most densely sampled grids computed for a single star to date. It consists of 10240 stellar tracks, all with atomic diffusion enabled \citep{Thoul94}, which is important for \methu \citep{ref:bond2013}. The tracks span the parameter space defined in Table~\ref{tab:GridParams}. The boundaries for the variable parameters $M$, $[\mathrm{Fe}/\mathrm{H}]$, and $[\alpha/\mathrm{Fe}]$ were chosen based on expectations from the literature (see Sects.~\ref{sec:intro} and \ref{sec:existingdata}). As \methu was anticipated to have a large age and a low mass -- as well as being a subgiant star -- we omitted the effects of convective overshooting and mass loss. The mixing length parameter, $\alpha_\mathrm{MLT}$ and the initial helium abundance, $Y_\mathrm{ini}$, were both treated as free parameters. This choice was made based on the considerations in \citet{LiY2024}, thus allowing for the possibility of resolving a tentative degeneracy between them in the obtained solutions. Stellar models were recorded in an evolutionary region around \methu, set by the value of $\Delta\nu$ lying within the bounds seen in Table~\ref{tab:GridParams}.

\subsection{Finding the best model}
\label{subsec:modelling}

The search for the solutions was carried out by the BAyesian STellar Algorithm (BASTA; \citealt{Aguirre22}), a Bayesian probabilistic pipeline to determine stellar properties. BASTA has previously been used for subgiant stars, in fact the frequency matching algorithm was developed specifically for subgiants \citep{Stokholm19}. The parameters used in the inference for \methu were $T_\mathrm{eff}$, $[\mathrm{Fe}/\mathrm{H}]$, parallax (see Table~\ref{tab:knownparams}), and (some of) the individual frequencies (Table~\ref{tab:freqs}). Including the parallax entailed sampling the distance to \methu through the Gaia $G_\mathrm{BP}$, $G_\mathrm{RP}$, and $G$ magnitudes. With BASTA we did not use the luminosity as a direct constraint during the inference, but rather formed the constraint through a combination of the parallax and apparent magnitudes (see Sect. 4.2.2 of \citealt{Aguirre22} for details on BASTA's distance module). The Gaia 5-parameter zero-point correction of \citet{Lindegren21} was applied to the parallax, and an uncertainty floor of $0.01$~mag was applied to all the magnitudes retrieved from \gaia to reflect the systematic offset between measured and synthetic magnitudes \citep{ref:riello2021}. Following \citet{ref:li2025} we excluded the interferometric radius as a direct constraint in the modelling to avoid any inconsistency between this value and the asteroseismic constraints.

Prior to the inference, the correction due to the Doppler-shift incurred by the radial velocity \citep{Davies14} was applied to the frequencies. For \methu the correction was $1+v_r/c = 0.9994317$ with $c$ being the speed of light and $v_r$ the radial velocity obtained from Gaia. For the model frequencies, the cubic term of the surface correction by \citet{Ball14} was used to shift them to the observed. We did not account for mode coupling in the surface term correction for \methu, although it can affect the resulting mass and age for subgiants \citep{ref:ong2021}, as \methu is less evolved than the limit where this becomes important \citep[$\Delta\nu \lesssim 30 \ \mu\mathrm{Hz}$,][]{ref:ong2021}. We also did not account for any potential metallicity effect on the surface correction, as has been suggested by \citet{ref:manchon2018}, who studied metallicities down to $[\mathrm{Fe}/\mathrm{H}] = -1$.


\section{Modelling results}
\label{sec:results}

\begin{table}
\caption{Stellar parameters obtained from the modelling.}
\label{tab:modparams}
\renewcommand{\arraystretch}{1.2}
\centering
\begin{tabular}{l c}
\hline\hline
Parameter & Value \\
\hline
    $T_\mathrm{eff} \ (K)$                      & $5818 \substack{+27\\-34}$ \\
    $[\mathrm{Fe}/\mathrm{H}] \ (\mathrm{dex})$ & $-2.23 \substack{+0.08\\-0.10}$ \\
    $\log(g) \ \log(\mathrm{cgs})$              & $3.679 \pm  0.002$ \\
    Age (Gyr)                                   & $14.2 \pm 0.4$ \\
    $M \ (\mathrm{M}_\odot)$                    & $0.75 \pm 0.01$ \\
    $[\mathrm{\alpha}/\mathrm{Fe}] \ (\mathrm{dex})$\tablefootmark{a} & $0.3 \pm 0.1$ \\
    $\nu_\mathrm{max} \ (\mu\mathrm{Hz})$       & $537.2 \substack{+2.9\\-1.8}$ \\
    $R \ (\mathrm{R}_\odot)$                    & $2.078 \substack{+0.012\\-0.011}$ \\
    $L \ (\mathrm{L}_\odot)$                    & $4.43 \substack{+0.10\\-0.12}$ \\
    $\alpha_\mathrm{MLT}$                       & $1.80 \substack{+0.05\\-0.06}$ \\
    $\mathrm{Y}_\mathrm{ini}$                   & $0.254 \substack{+0.004\\-0.005}$ \\
\hline
\end{tabular}
\tablefoot{The quoted values are the median of the posterior distribution along with the $68$-percentiles. The reported metallicity, $[\mathrm{Fe}/\mathrm{H}]$, and $\alpha$-abundance, $[\mathrm{\alpha}/\mathrm{Fe}]$, were derived from the surface values at the end of evolution.
\tablefoottext{a}{Due to crude sampling, see Sec.~\ref{subsec:modelgrid} for details.}
}
\end{table}

The results of the modelling using the six radial modes, 11 dipolar modes (from Table~\ref{tab:freqs}), Gaia parallax and colours, as well as the effective temperature and metallicity (see Table~\ref{tab:knownparams}) as constraints can be found in Table~\ref{tab:modparams}. The values quoted here are the medians of the respective posteriors, with uncertainties from the 16th and 84th percentiles. The full posterior distributions are shown in Fig.~\ref{fig:corner}. For completeness, we also performed the model inference using the 2MASS colours ($J$, $H$ and $K$), which yielded the exact same best-fitting model and indistinguishable posteriors.

\begin{figure}
   \centering
   \includegraphics[width=\hsize]{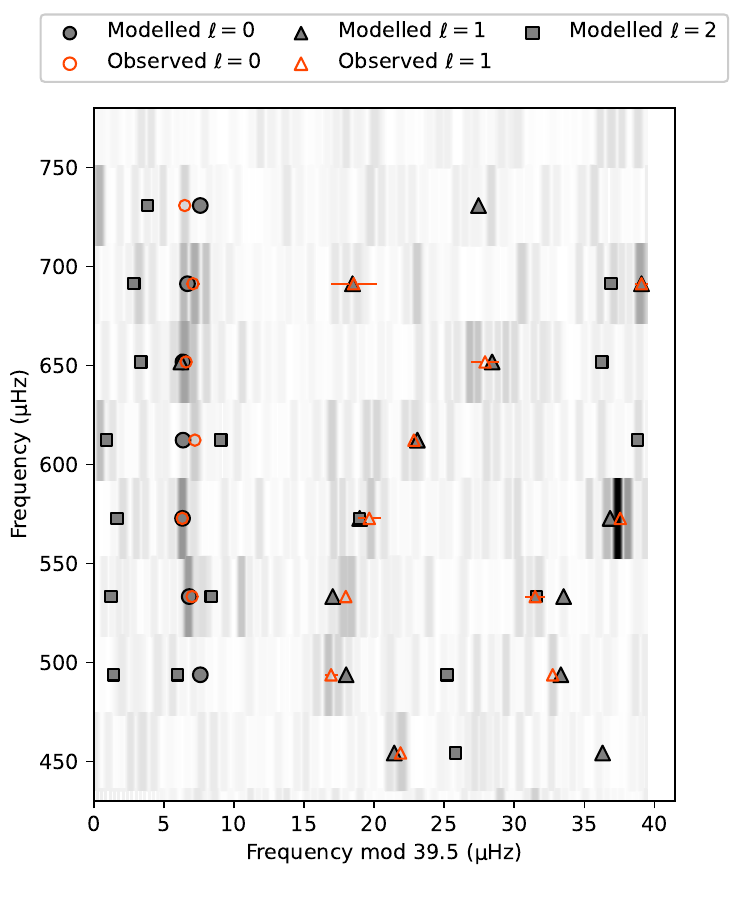}
      \caption{{\'E}chelle diagram showing the observed (orange) and modelled (grey) mode frequencies for $\ell=0$ (circles), $\ell=1$ (triangles), and modelled $\ell=2$ (squares). The observed modes are plotted with their $1\sigma$ uncertainties in the horizontal direction if these are larger than the symbol used. The straight ridge of the radial modes and the mixed-mode behaviour of the dipolar modes are evident. Note that model modes without an observed counterpart are also plotted.
              }
    \label{fig:echelle}
\end{figure}

The surface-corrected frequencies from the best-fitting model can be seen plotted with the (Doppler-shifted) observed frequencies in an {\'e}chelle diagram in Fig.~\ref{fig:echelle}. Here, $\ell=1$ mixed-modes are dominating the visual impression of the diagram, while the $\ell=0$ ridge is evident on the left side of the diagram. It can be seen that overall there is a close agreement between the observed and modelled modes. The majority of the mixed modes are well-represented by the best-fitting model, which is noteworthy as their frequency changes rapidly during this phase of the evolution \citep[see, e.g.][]{Stokholm19}. Encouragingly, a few of the $l=2$ model frequencies (which were not used in the fitting) lie close to peaks in the observed power spectrum.

Our final mass of $0.75 \pm 0.01 \ \mathrm{M}_\odot$ is consistent with the literature, for instance, it agrees well with the value of $0.77 \pm 0.03 \ \mathrm{M}_\odot$ found in the interferometric analysis by \citet{ref:karovicova2020}. Our value is also in agreement with the values found in the recent analysis using custom abundances by \citet{ref:guillaume2024}.

The radius inferred from the asteroseismic modelling agrees with the interferometric radius at the $1.5\sigma$ level. Thereby, the agreement is better than what was found for the K-dwarf HD~219134 by \citet{ref:li2025} ($4\sigma$), and similar to that found by \citet{Stokholm19} for the sub-giant HR~7322 ($1.5\sigma$). We note that, in all three cases, the radius from the seismic modelling is smaller than the one from interferometry. Our inferred stellar radius is ${\sim}3\%$ smaller than the one from interferometry, which is in broad agreement with the finding by \citet{ref:huber2017} that seismic radii are underestimated by ${\sim}5\%$.

Re-deriving the luminosity from \citet{ref:karovicova2020} using the updated Gaia parallax yields $L = 4.633 \pm 0.037 \ \mathrm{L}_\odot$, which agree to $1.7\sigma$ with our modelled luminosity (see Table~\ref{tab:modparams}). This difference is reasonable given that we neither used the luminosity as a direct constraint in the model inference (see Sect~\ref{subsec:modelling}) nor included the interferometric radius as a constraint in the modelling.


\subsection{The age of HD 140283}
\label{subsec:age}

Our seismically constrained age ($14.2 \pm 0.4$\,Gyr) is higher by $0.8\sigma$ than the estimated age of the Universe of $13.787 \pm 0.020 \ \mathrm{Gyr}$ \citep{ref:planck2020}. Thereby, from our analysis, the age of \methu is not at odds with the age of the Universe, which has sometimes been the case in the past (refer to Fig.~\ref{fig:age_lit} or Table~\ref{tab:app_ages}). It can be concluded that \methu is an old star that was born early in the history of the Universe. 
Here, it should also be noted that the uncertainties are those quoted by BASTA\footnote{
BASTA takes correlations between the different parameters into account, which in general provides more realistic uncertainties. Furthermore, we have changed some of the model physics in our modelling by allowing, for instance, the mixing length parameter and the initial helium abundance to vary freely; the latter as an alternative to using a galactic enrichment law. However, we have not varied, for example, the prescriptions for the convection or the surface correction, or the weighting scheme for the frequencies, which impacts the uncertainties too \citep{ref:cunha2021}.
}, not accounting for all systematic effects. As suggested by, for instance, \citet{ref:tayar2022}, realistic age uncertainties for sub-giants are higher, for \methu perhaps of the order of 10\% (estimated from their Fig.~6).

The mixing length parameter favoured by our analysis $\alpha_\mathrm{MLT} = 1.80 \pm 0.05$ is consistent with the value GARSTEC determines for the Sun. Thus, in our analysis, there is no evidence for less efficient mixing in \methu. This is interesting because it was found by \citet{ref:guillaume2024} that a reduced mixing would lower the age of \methu, although---in agreement with our result---they find no direct evidence of a lower-than-solar mixing.

Other physical effects such as diffusion and opacities may also affect the age \citep{ref:tang2021, ref:guillaume2024}. The same is true for using custom abundances rather than the solar-scaled mixture \citep{ref:guillaume2024} that we have adopted here due to limitations in the availability of custom opacity tables in GARSTEC. However, for testing purposes, we produced a small stellar grid of $N=1096$ tracks with an unrealistically inflated $[\alpha/\mathrm{Fe}]=0.6$ dex. This artificial enhancement increases the oxygen abundance by the same amount, while simultaneously applying identical changes to the remaining alpha elements. The parameters used for the inference were identical to those outlined in Sec.~\ref{subsec:modelling}. The effect on the obtained age was small, recovering a similar age of $14.1^{+0.1}_{-0.3}$ Gyr. However, the posterior in age has become bimodal and indicates a tentative solution at even lower age, in line with the expectations from \citet{ref:guillaume2024} that an increased oxygen abundance results in a lower age for \methu.


\subsection{Impact of using different constraints during the model inference}
\label{subsec:other_constraints}

As mentioned at the beginning of this section, the results described above are based on model inference using the radial and dipolar modes listed in Table~\ref{tab:freqs}, and excluding the interferometric radius. We have also carried out the model inference using:
\begin{enumerate}
    \item\label{item:radial} All radial modes and the two most prominent $\ell=1$ modes (at $590.89$ and $710.98 \ \mu$Hz). We did this to gauge the impact of including potentially misidentified $\ell=1$ modes.
    \item\label{item:l2} All modes present in Table~\ref{tab:freqs} (that is, including the one $\ell=2$ mode). As mentioned in Sec.~\ref{subsec:freqs}, we omitted $\ell=2$ modes in the reported model inference; however, as $\ell=2$ modes can, in general, help to constrain the age (through the small spacing), we included it to test the effect on the age of \methu.
    \item\label{item:Rint} All radial modes, all $\ell=1$ modes, and the interferometric radius (see Table~\ref{tab:knownparams}). As the interferometric radius is available, we tested the impact on the results of adding it as a constraint in the model inference.
    \item\label{item:inflated} All radial modes, all $\ell=1$ modes, and the interferometric radius with inflated uncertainties. It is suggested by \citet{ref:tayar2022} that typical limits of $4\% \pm 1\%$ apply to measurements of interferometric radii. This, and the fact that the constraint bands from the Gaia colours, frequencies, effective temperature, and interferometric radius did not overlap in a Kiel diagram (see Fig.~\ref{fig:app_kiel}), motivated us to do a modelling round where we inflated the uncertainties on the interferometric radius to 5\%.
\end{enumerate}

In all four cases, there were minor shifts in some of the parameters and their uncertainties, but the outcomes of the model inference were consistent to within $1\sigma$ with the result presented above (see Table~\ref{tab:modparams}). It is perhaps noteworthy that introducing a quadrupolar mode does not affect the age or its uncertainty, but this is hardly surprising based on the results of \citet{ref:white2011} and \citet{ref:ong2025}, and on the fact that the $\ell=1$ mixed modes---which dominate our model inference---already carry information on the evolutionary state \citep[e.g.][]{ref:campante2023}.

The largest change to the inferred radius came from the inclusion of the interferometric radius (case~\ref{item:Rint}), followed by the case where only two dipolar modes were included in the fitting (case~\ref{item:radial}). However, in both cases the change was less than $0.3\%$. The remaining two cases led to virtually no change in $R$, from which we conclude that the employed uncertainty on the interferometric radius only had a small effect on the modelled radius.


\section{Discussion}
\label{sec:discussion}

\subsection{Deviation from the $\nu_\mathrm{max}$ scaling relation}
\label{subsec:fnumax}

Asteroseismology of metal-poor stars not only improves our ability to date old stars such as \methu, but also provides insight into the validity and limitations of the widely used seismic scaling relations. 
If the temperature, radius, and mass of a star are known, then the \numax\ scaling relation $\nu_\mathrm{max} \propto \frac{M}{R^2 \sqrt{T_\mathrm{eff}}}$ \citep{ref:brown1991, ref:kjeldsen1995, Belkacem2011} can be used to predict the value of \numax. Several versions of this scaling relation exist, and it has been extensively validated  \citep{ref:silvaaguirre2012, ref:gaulme2016, ref:huber2017, ref:brogaard2018, ref:kallinger2018, ref:sahlholdt2018, ref:hall2019, ref:khan2019, ref:zinn2019, Hekker2020, ref:benbakoura2021, ref:li2021, LiY2024, ref:valle2025}. \citet{ref:viani2017} suggested the addition of a mean molecular weight term. 
Using recent 3D hydrodynamical simulations of a star similar to the Sun (with metallicities ranging from $-3$ to $0.5$), \citet{ref:zhou2024} did not find a change of $\nu_\mathrm{max}$ with metallicity. However, this could be due to an opposing effect from the Mach number of the simulations \citep{ref:zhou2024} or differences between the simulated solar-type star and red giants.

BASTA uses the version of the $\nu_\mathrm{max}$-scaling relation from \citet{ref:stello2008}, involving the stellar luminosity:
\begin{equation}
    \label{eq:numax_L}
    \nu_\mathrm{max} = \nu_{\mathrm{max},\odot} \frac{
                            (M / \mathrm{M}_\odot) (T_\mathrm{eff} / \mathrm{T}_{\mathrm{eff},\odot})^{3.5}
                            }{
                            L / \mathrm{L}_\odot
                            } \ .
\end{equation}
The reference value of the solar $\nu_\mathrm{max}$ is the same as that used in pySYD, namely $\nu_{\mathrm{max},\odot} = 3090 \ \mu$Hz \citep{ref:huber2011, Aguirre22}, while the other solar values are those used in BaSTI \citep{ref:hidalgo2018}: $\mathrm{T}_{\mathrm{eff},\odot} = 5777 \ \mathrm{K}$, $\mathrm{M}_\odot = 1.9891 \cdot 10^{33} \ \mathrm{g}$ and $\mathrm{L}_\odot = 3.842 \cdot 10^{33} \ \mathrm{erg} \cdot \mathrm{s}^{-1}$.

\begin{figure}
   \centering
   \includegraphics[width=\hsize]{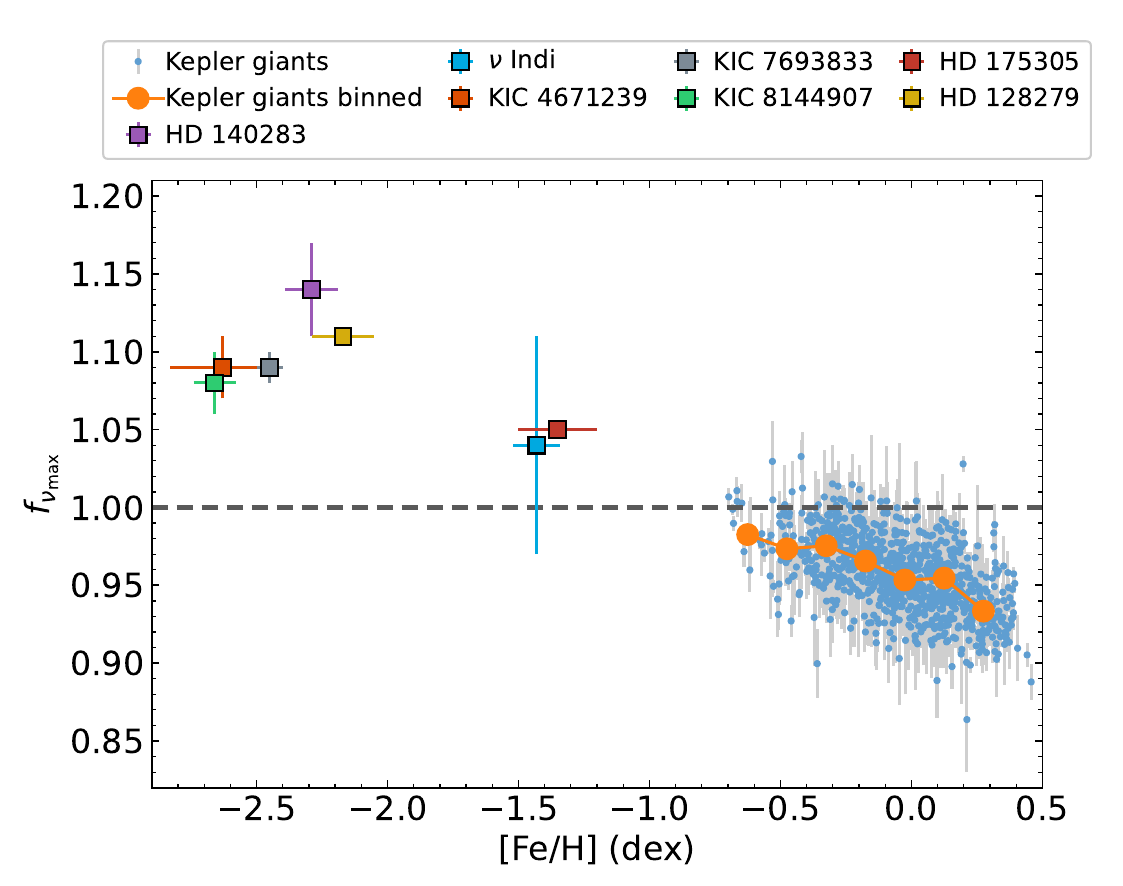}
      \caption{$f_{\nu_\mathrm{max}}$ as a function of metallicity for Kepler red giants \citep{LiY2024}, the low metallicity red-giants KIC~8144907 \citep{ref:huber2024}, KIC 7693833 and KIC 4671239  \citep{Larsen25}, and HD 175305 and HD 128279 \citep{ref:lindsay2025}, as well as the subgiants $\nu$~Indi \citep{ref:chaplin2020} and \methu.
              }
   \label{fig:fnumax}
\end{figure}

From our model inference we have the posterior distribution for the model-$\nu_\mathrm{max}$, $\left(\nu_\mathrm{max}\right)_\mathrm{mod}$, computed from Eq.~\ref{eq:numax_L} (see Table~\ref{tab:modparams}): $\left(\nu_\mathrm{max}\right)_\mathrm{mod} = 537.2 \substack{+2.9\\-1.8} \ \mu\mathrm{Hz}$. When comparing this value to the observed one, and symmetrising the uncertainties on the model-$\nu_\mathrm{max}$, we can determine the correction factor $f_{\nu_\mathrm{max}}$ \citep{ref:sharma2016}; that is the factor that needs to be applied to bring the observed and modelled $\nu_\mathrm{max}$ into agreement:
\begin{equation}
    \label{eq:fnumax}
    \left( \nu_\mathrm{max} \right)_\mathrm{obs} = f_{\nu_\mathrm{max}} \left(  \nu_\mathrm{max} \right)_\mathrm{mod} \ .
\end{equation}
From this, we find $f_{\nu_\mathrm{max}} = 1.14 \pm 0.03$ for \methu. We can compare this to other metal-poor stars in the literature; KIC4671239 and KIC7693833 \citep{Larsen25}, $\nu$~Indi \citep[][we use $\left(\nu_\mathrm{max} \right)_\mathrm{obs} = 342 \pm 3 \ \mu\mathrm{Hz}$]{ref:chaplin2020}, KIC~8144907 \citep{ref:huber2024}, and HD 175305 and HD 128279 \citep{ref:lindsay2025}. Their $f_{\nu_\mathrm{max}}$ values, as well as that of \methu, can be seen as a function of metallicity in Fig.~\ref{fig:fnumax}. Also shown are the $f_{\nu_\mathrm{max}}$ results of \citet{LiY2024}, which we have re-derived using an Eddington grey atmosphere, as is used in the models for \methu, and employing values of the mixing length parameter and initial helium abundance that are similar to those found for \methu. We see that the seven metal-poor stars extend the trend that is apparent from the higher-metallicity giant-star sample of \citet{LiY2024}, confirming an increasing discrepancy between the model- and observed \numax as the stellar metallicity is reduced. This was also noted by \citet{TLi2022}, \citet{ref:huber2024} and \citet{Larsen25}. The sample of the seven metal-poor stars comprises two sub-giants and five red giants, with \methu having the largest $f_{\nu_\mathrm{max}}$. It would be interesting to add more metal-poor stars to this plot to validate the trend and to investigate whether it relates to the stellar evolutionary state (T. Li et al, in prep.; Sreenivas et al., in prep.).

\subsection{Galactic origin}
\label{subsec:galorigin}

\begin{figure}
   \centering
   \includegraphics[width=\columnwidth]{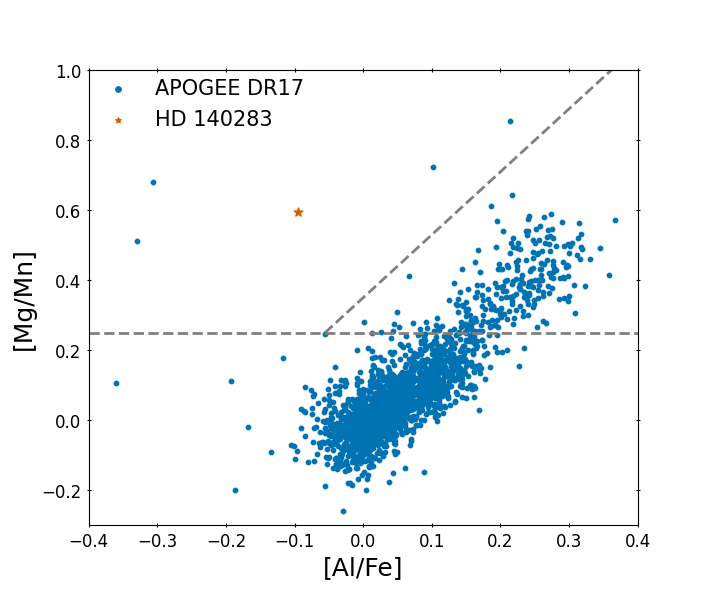}
      \caption{Chemical origin plot showing the $[\mathrm{Mg}/\mathrm{Mn}]$ and $[\mathrm{Al}/\mathrm{Fe}]$ abundance ratios for \methu (orange) compared to those for the stars of APOGEE DR17 (blue). The grey lines shows areas of this space dominated by different stellar populations, with the lower being the Galactic thin disk, the upper right the Galactic thick disk, and the upper left stars born ex situ \citep{horta2021}.
              }
    \label{fig:gal_chem}
\end{figure}

To investigate the Galactic origin of \methu, we used the abundances by \citet{ref:amarsi2019} and \citet{ref:amarsi2022} \citep[see][for details]{ref:guillaume2024} to compute the $[\mathrm{Mg}/\mathrm{Mn}]$ and $[\mathrm{Al}/\mathrm{Fe}]$ abundance ratios. We used these to study the [Mg/Mn]-[Al/Fe] plane in order to assess whether \methu is chemically most similar to the in situ or ex situ populations of stars in the Milky Way. 
The choice of plane was motivated by \citet{hawkins2015} and introduced in this context by \citet{das2020}, and it neatly separates the thin- and thick-disk stars born in situ from accreted stars born ex situ.
This plane has been successfully used in a number of studies \citep[see e.g.][]{horta2021,borre2022} in order to study Galactic substructure and disentangle stellar remnants of other systems from stars born in the Milky Way potential.

From Fig.~\ref{fig:gal_chem}, we see that the location of \methu in this plane is different from the vast majority of thin- and thick disk stars in APOGEE DR17 \citep{ref:abdurrouf2022}. The relatively high value of $[\mathrm{Mg}/\mathrm{Mn}]$, combined with the comparatively low value of $[\mathrm{Al}/\mathrm{Fe}]$, indicates that \methu was born ex situ. 

\begin{figure*}
   \includegraphics[width=0.45\textwidth]{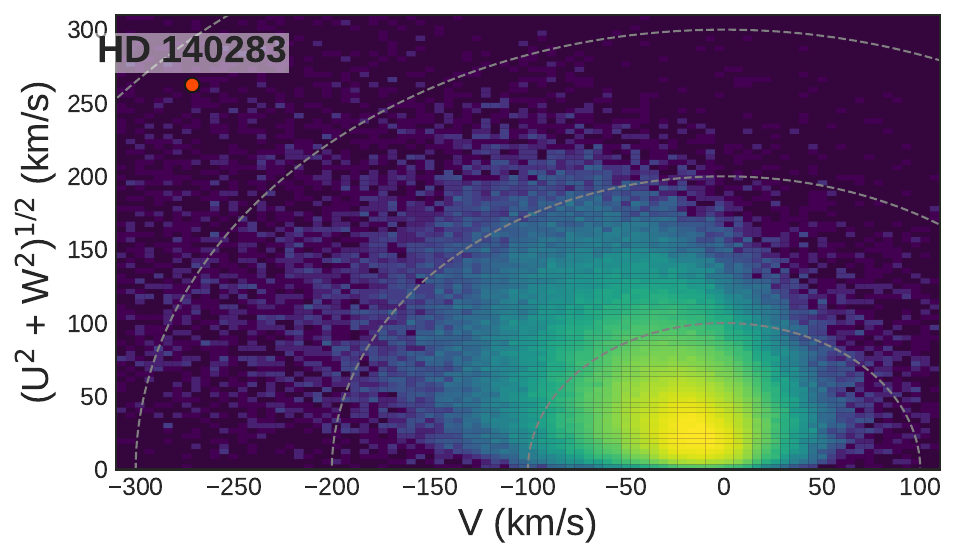}
    \hfill
    \includegraphics[width=0.45\textwidth]{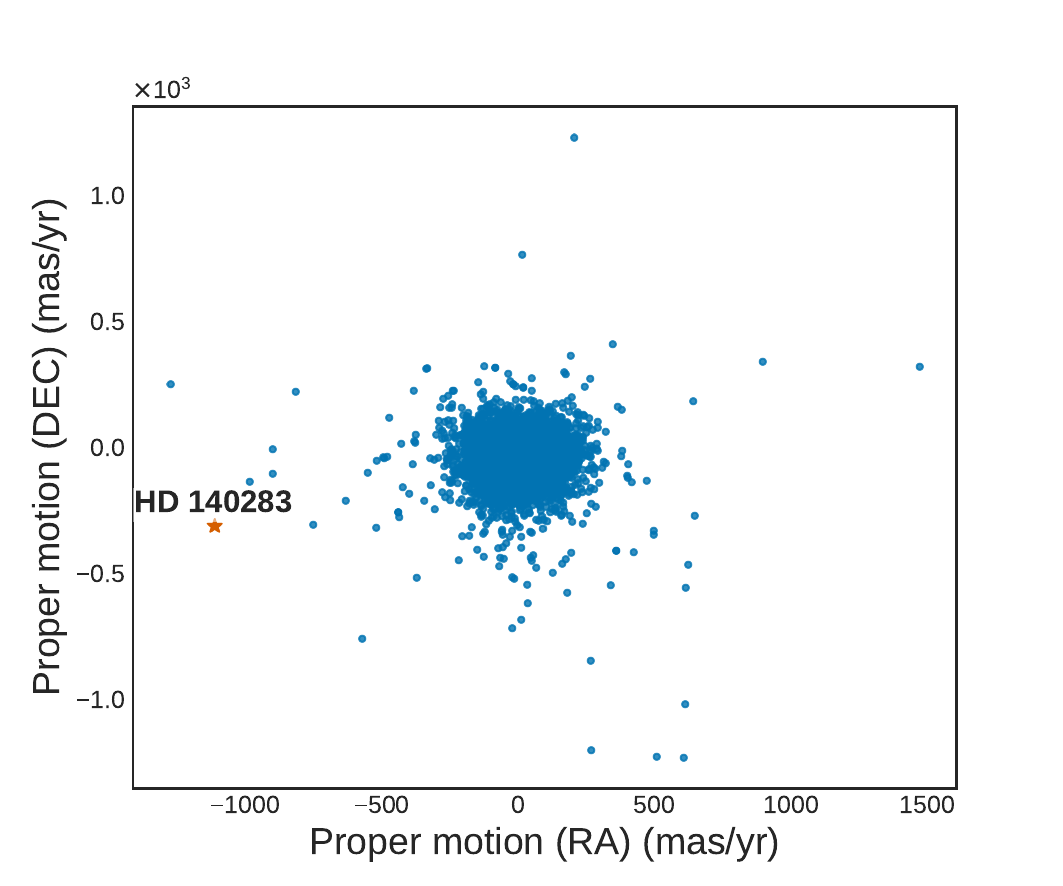}
    \\[1ex]
    \includegraphics[width=0.45\textwidth]{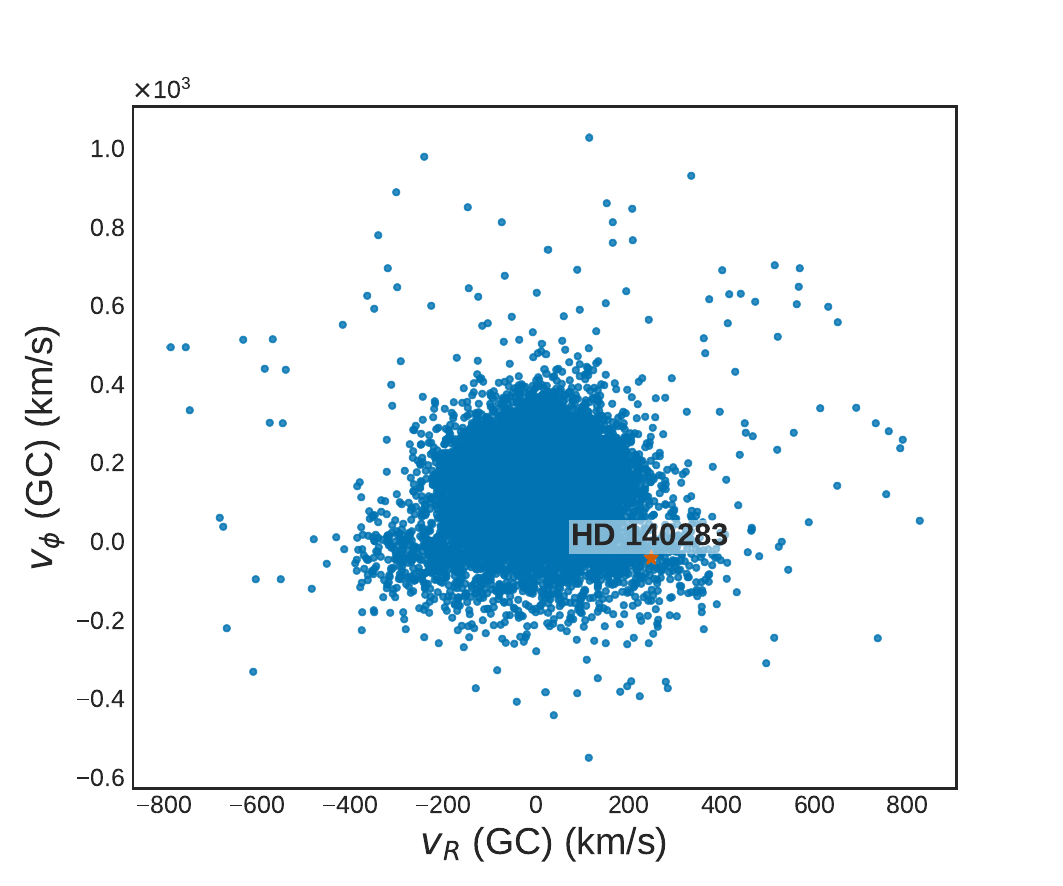}
    \hfill
    \includegraphics[width=0.45\textwidth]{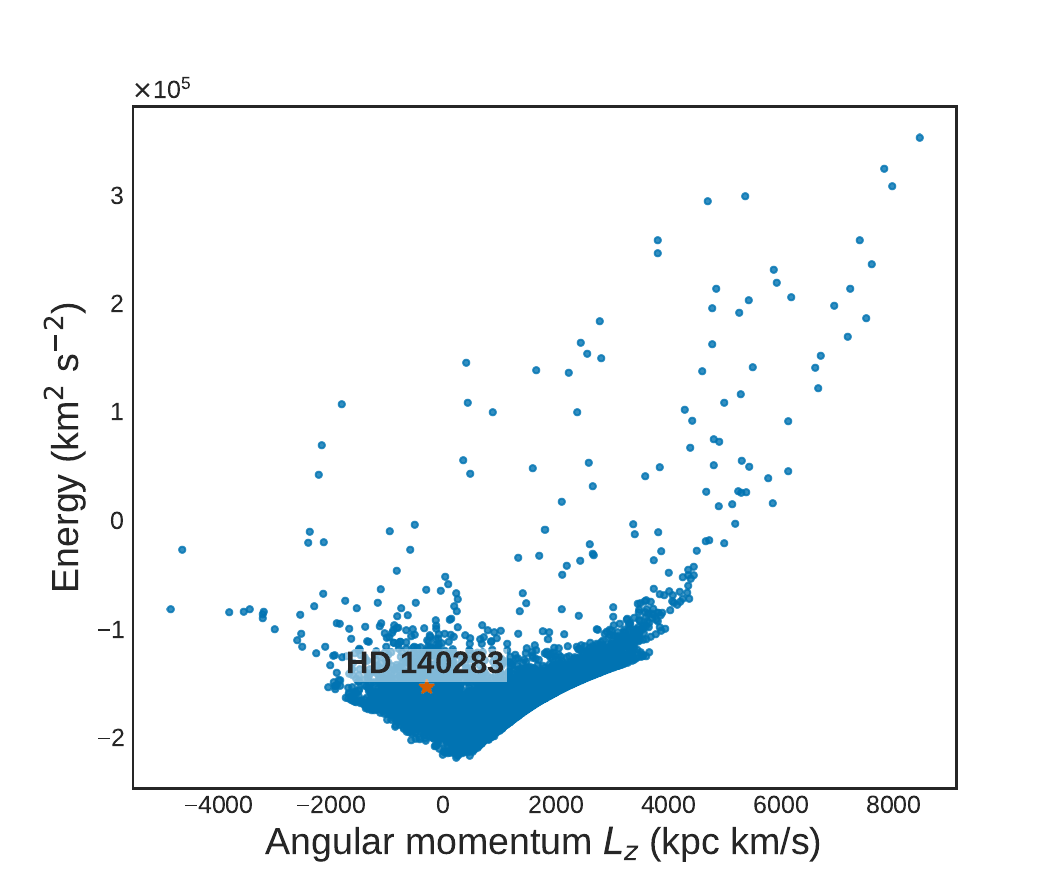}
      \caption{Plots of the Galactic orbital motion of \methu compared single stars from \gaia DR3 with reliable astrometric data and available line-of-sight velocities (\texttt{astrometric\_params\_solved = 95}, \texttt{rv\_nb\_transits > 0}, \texttt{ruwe < 1.4}, and \texttt{non\_single\_star = 0}). \textit{Top left:} Toomre diagram showing the location of \methu in heliocentric cartesian velocity-space. 
      \textit{Top right:} Plot of proper motion space. \textit{Bottom left:} Plot showing the Galactocentric cylindrical coordinates $V_{\phi}$ vs $V_R$.
      \textit{Bottom right:} Energy-angular momentum space on the energy-angular momentum plane. 
              }
      \label{fig:combined_orbits}
\end{figure*}
Another check on the Galactic origin can come from evaluating the orbit of \methu in the Milky Way.
\methu stands out as a clear outlier in proper motion space compared to the bulk of stars in the \gaia DR3 catalogue, as can be seen in Fig.~\ref{fig:combined_orbits}. We computed the Galactic orbit of \methu using the 5D astrometric information and line-of-sight velocity from \gaia DR3 \citep{ref:gaia2016, ref:gaiadr3_2023}, applying the parallax zero-point correction from \citet{Lindegren21}.
We computed the Galactic orbital properties using \texttt{galpy} \citep{bovy2015}, adopting the axisymmetric gravitational potential from \citet{mcmillan2017} as a model for the Milky Way. For the solar position and velocity, we assumed $(X_{\odot}, Y_{\odot}, Z_{\odot}) = (8.2, 0, 0.0208)$~\si{\kilo pc}, with a circular velocity of \SI{240}{\kilo\metre\per\second} \citep{schonrich2010,bovy2015,bennett2019,gravity2019}.

The resulting orbit of \methu, shown in Fig.~\ref{fig:combined_orbits}, is highly eccentric and retrograde. These characteristics are consistent with a Galactic halo origin. In both velocity space and energy–angular momentum space, \methu occupies the region typically associated with Gaia-Enceladus \citep{helmi2018,belokurov2018,feuillet2020}.

Taken together---the asteroseismic age, chemical abundances, and orbital dynamics---all lines of evidence support the conclusion that \methu is a halo star. It was presumably accreted from an external galaxy, most plausibly Gaia-Enceladus, during its merger with the Milky Way $8$-$11$~Gyr ago.

\section{Conclusion}
\label{sec:conclusion}

In this work we presented the detection of stellar oscillations in the bright metal-poor subgiant \methu, also known as the Methuselah star. These were used in combination with the parallax, \gaia colours, metallicity, effective temperature, and alpha-element enhancement as constraints on the modelling of \methu using BASTA. We inferred a mass of $M = 0.75 \pm 0.01 \ \mathrm{M}_\odot$ and a radius of $R = 2.078 \substack{+0.012\\-0.011} \ \mathrm{R}_\odot$. This radius is $1.5\sigma$ smaller than the interferometric radius computed from the angular diameter by \citet{ref:karovicova2020}, supporting the finding that seismic radii are underestimated \citep{ref:huber2017}.

The age determined from the modelling was $t = 14.2 \pm 0.4 \ \mathrm{Gyr}$ (internal uncertainties only), in agreement with the age of the Universe within $1\sigma$. In line with the results of \citet{ref:guillaume2024}, we found no evidence for the mixing-length parameter to be different to the solar value, which would otherwise be a path to lowering the determined age (see Sec.~\ref{subsec:age}). Interestingly, we found indications in both the kinematics and the chemical abundances that \methu originated outside our Milky Way, with the kinematics suggesting that it may be a member of Gaia-Enceladus.

We determined that the frequency of maximum oscillation power is larger than expected from the standard scaling relation, by a factor of $f_{\nu_\mathrm{max}} = 1.14 \pm 0.03$. This result supports the trend found in, for instance, \citet{LiY2024}, that the discrepancy between the modelled and observed $\nu_\mathrm{max}$ increase as we move to more metal-poor stars. It will be interesting to investigate this trend for other metal-poor stars in different evolutionary stages.

TESS has very recently concluded observing \methu for an additional sector, doubling the amount of available 20~second cadence data. Performing a new asteroseismic analysis including this new data, could lead to the detection of additional oscillation modes, providing extra constraints for the model inference. Another, more important, update on the asteroseismic analysis provided here will come from the inclusion of custom abundances and associated opacities as it was shown by \citet{ref:guillaume2024} that this can impact the results of the modelling, notably the age.

Further scrutiny of \methu, the ‘Methuselah’ star, will not only advance our understanding of this particular star, but also of oscillations in very metal-poor stars in general. Individual oscillations have only been detected in very few stars of this low metallicity (see Sec.~\ref{subsec:fnumax}). \methu represents, to our knowledge, the highest-$\nu_\mathrm{max}$ metal-poor star where a detailed asteroseismic modelling, including mixed modes, has been possible to date.

Furthermore, the low metallicity of \methu combined with its high oxygen and nitrogen abundance makes it an ideal test-bed for the mixing properties acting in Pop III stars, allowing to test the validity of the predictions of fast rotating models of the primordial stellar generation \citep[see e.g.][]{ref:chiappini2011,ref:Tsiatsiou2024} from which \methu was formed.


\begin{acknowledgements}
      The authors thank Mikkel S. Lund for providing the $\nu_\mathrm{max}$ value for $\nu$~Indi. 
      This work was supported by a research grant (42101) from VILLUM FONDEN as well as The Independent Research Fund Denmark’s Inge Lehmann program (grant agreement No. 1131-00014B). Funding for the Stellar Astrophysics Centre was provided by The Danish National Research Foundation (grant agreement No. DNRF106).
      T.R.B. is supported by an Australian Research Council Laureate Fellowship (FL220100117).
      G.B. acknowledges fundings from the Fonds National de la Recherche Scientifique (FNRS) as a postdoctoral researcher.
      MBN acknowledge the support from the UK Space Agency.
      This paper includes data collected by the TESS mission, which are publicly available from the Mikulski Archive for Space Telescopes (MAST). Funding for the TESS mission is provided by NASA’s Science Mission directorate.
      This research has made use of the SIMBAD database, operated at CDS, Strasbourg, France.
      This work has made use of data from the European Space Agency (ESA) mission Gaia (https://www.cosmos.esa.int/gaia), processed by the Gaia Data Processing and Analysis Consortium (DPAC, https://www.cosmos.esa.int/web/gaia/dpac/consortium). Funding for the DPAC has been provided by national institutions, in particular, the institutions participating in the Gaia Multilateral Agreement.
      This research has made use of the Exoplanet Follow-up Observation Program (ExoFOP; DOI: 10.26134/ExoFOP5) website, which is operated by the California Institute of Technology, under contract with the National Aeronautics and Space Administration under the Exoplanet Exploration Program.
      Some observations in this paper made use of the High-Resolution Imaging instrument Zorro. Zorro was funded by the NASA Exoplanet Exploration Program and built at the NASA Ames Research Center by Steve B. Howell, Nic Scott, Elliott P. Horch, and Emmett Quigley. Zorro is mounted on one of the 8.1-m telescopes of the international Gemini Observatory (Gemini South), a program of NSF NOIRLab, which is managed by the Association of Universities for Research in Astronomy (AURA) under a cooperative agreement with the U.S. National Science Foundation, on behalf of the Gemini partnership: the National Science Foundation (United States), National Research Council (Canada), Agencia Nacional de Investigaci{\'o}n y Desarrollo (Chile), Ministerio de Ciencia, Tecnolog{\'i}a e Innovaci{\'o}n (Argentina), Minist{\'e}rio da Ci{\^e}ncia, Tecnologia, Inova{\c{c}}{\~o}es e Comunica{\c{c}}{\~o}es (Brazil), and Korea Astronomy and Space Science Institute (Republic of Korea).
\end{acknowledgements}

\bibliographystyle{aa}
\bibliography{hd140283_bib}

\begin{appendix}

\section{Literature ages for HD~140283}
\label{sec:app_ages}

The ages and masses that were included in Fig.~\ref{fig:age_lit} can be found in Table~\ref{tab:app_ages} along with their associated references.

\begin{table}[h!]
\caption{Ages and masses of \methu\ as found in the literature}
\label{tab:app_ages}
\centering
\begin{tabular}{l l l l c}
\hline\hline            
Abbreviation & Age (Gyr) & Mass ($\mathrm{M}_\odot$) & Reference & Comment \\ 
\hline                    
   V00  & $\leq 16$                        & & \citet{ref:vandenberg2000} &  \\ 
   I02  & $11.38$                           & & \citet{ref:ibukiyama2002} & \\
   V02  & $13.5 \pm 1.5$                   & & \citet{ref:vandenberg2002} &  \\
   C11p & $13.42 \substack{+0.32\\-0.55} $ & $0.80 \pm 0.01$ & \citet{ref:casagrande2011} & Padova \\
   C11b & $13.61 \substack{+0.25\\-0.54} $ & $0.78 \pm 0.01$ & \citet{ref:casagrande2011} & BaSTI \\
   B13  & $14.46 \pm 0.31$                 & & \citet{ref:bond2013}       &  \\
   B14  & $14.7 \substack{-0.9\\-6.6}$     &$0.78 \substack{+0.16\\+0.01}$ & \citet{ref:bensby2014}     & Plotted as ranges $[8.1 - 13.8]$ and $[0.79, 0.94]$ \\
   V14  & $14.27 \pm 0.38$                 & & \citet{ref:vandenberg2014} & \\
   C15  & $13.7 \pm 0.7$                   & $0.78 \pm 0.01$ & \citet{ref:creevey2015}    & $A_\mathrm{V} = 0$ results \\
   J18  & $ [12.5, 14.9] $                 & $[0.75, 0.79]$ & \citet{ref:joyce2018}      & Range of ages and masses given in their Table 2  \\
   S19  & $\geq 12$                        & & \citet{ref:sahlholdt2019}  &  \\
   C20  & $14.42 \substack{+2.96\\-1.28}$  & $0.73 \substack{+0.04\\-0.02}$ & \citet{ref:chen2020}       &  \\
   K20 &                                   & $0.77 \pm 0.03$ & \citet{ref:karovicova2020} & \\
   T21  & $12.0 \pm 0.5$                   & $0.81 \pm 0.05$ & \citet{ref:tang2021}       &  \\
   P22b & $14.41 \pm 0.65$                 & &\citet{ref:plotnikova2022} & BaSTI average \\
   P22p & $14.69 \pm 1.03$                 & &\citet{ref:plotnikova2022} & Padova average \\
   G24s & $13.08 \pm 0.85$                 & $0.772 \pm 0.015$ & \citet{ref:guillaume2024}  & Custom abundances, solar scaled mixing \\
   G24r & $12.60 \pm 0.88$                 & $0.780 \pm 0.015$ & \citet{ref:guillaume2024}  & Custom abundances, 9\% reduced mixing \\
\hline                               
\end{tabular}
\end{table}

\section{Graphical output from the modelling}
\label{sec:app_modellingoutput}

\begin{figure}[h!]
   \centering
   \includegraphics[width=\hsize]{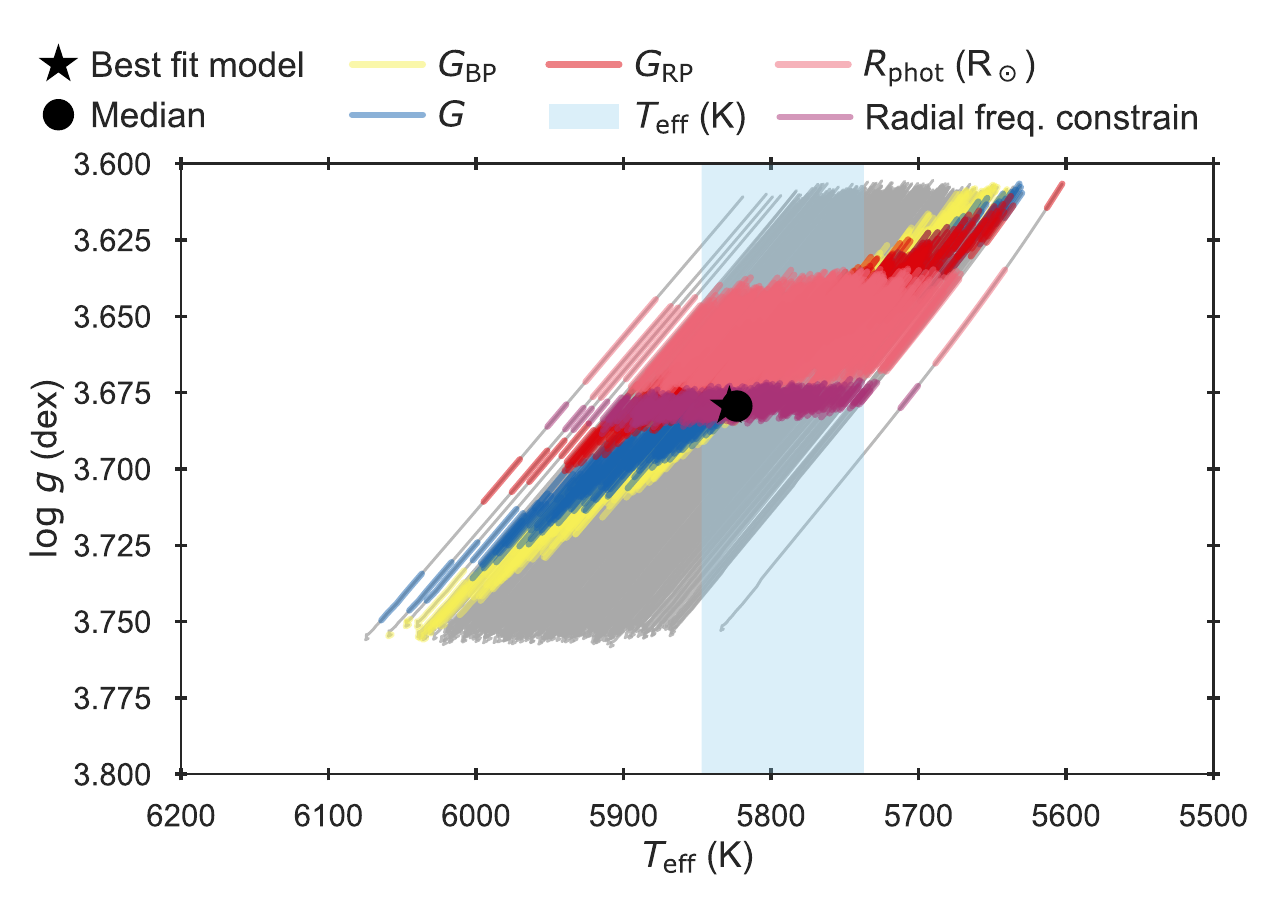}
      \caption{Kiel diagram for \methu showing the $1\sigma$ constraint bands from the Gaia colours (yellow, dark blue, and red), the frequencies ($\ell=0$ and $\ell=1$, purple), the effective temperature (light blue), and the interferometric radius (pink). The best-fitting model (star) and the median values (circle) are indicated.
              }
         \label{fig:app_kiel}
\end{figure}

When including the radius derived from interferometry as a constraint on the model inference (see case 3 in Sect.~\ref{subsec:other_constraints}), the Kiel diagram is as shown in Fig.~\ref{fig:app_kiel}. It can be seen that the constraint band from the frequencies and that from the radius do not, in general, overlap.

\begin{figure*}
   \resizebox{\hsize}{!}{\includegraphics{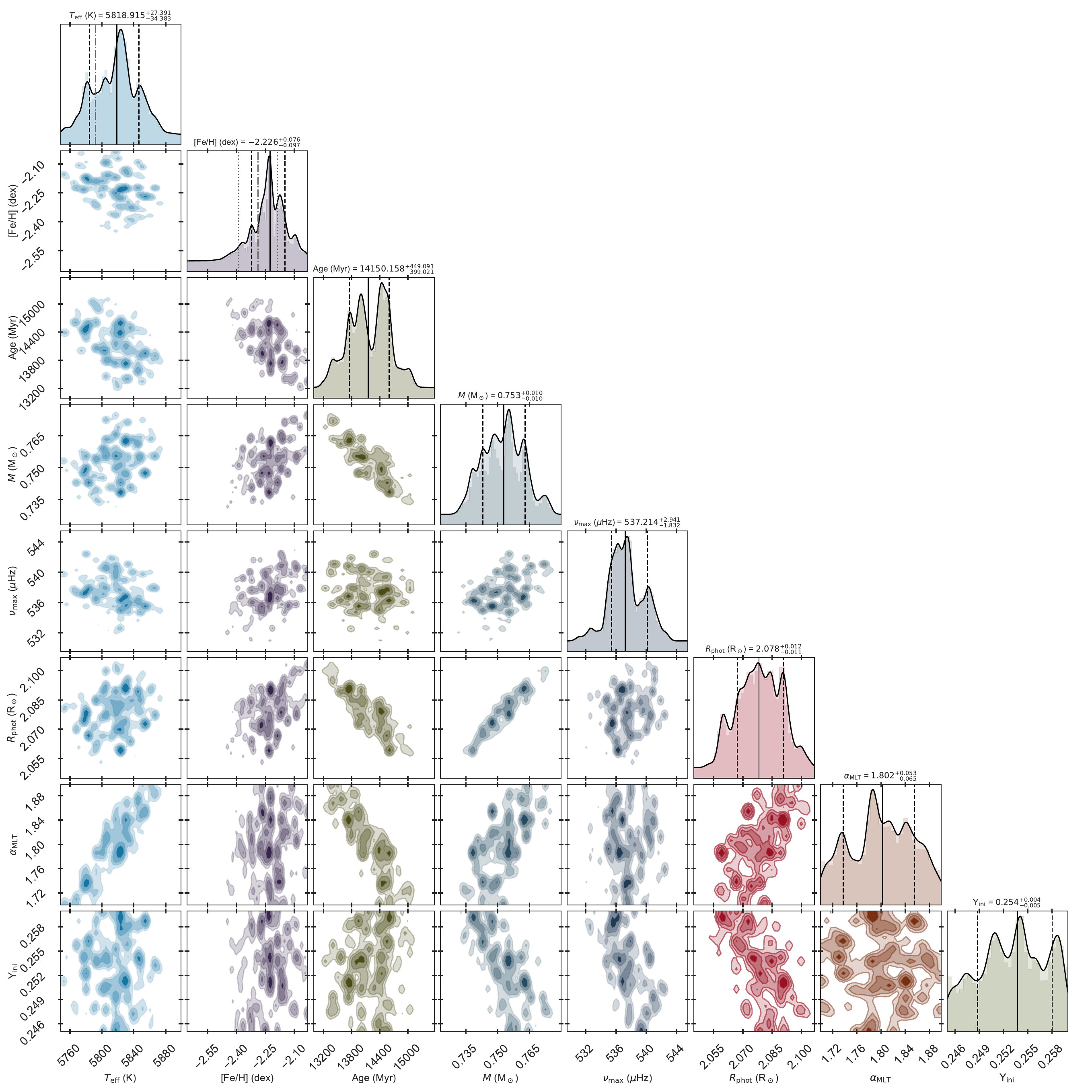}}
      \caption{Corner plot showing posterior distributions for the modelling using as constraints the six radial- and 11 dipolar modes listed in Table~\ref{tab:freqs} along with the parallax, Gaia magnitudes, effective temperature, and metallicity (see Table~\ref{tab:knownparams}).
              }
      \label{fig:corner}
\end{figure*}

Figure~\ref{fig:corner} shows the posterior distributions of key inferred model parameters.

\end{appendix}

\end{document}